\documentclass[12pt,onecolumn]{emulateapj}

\usepackage{epsf}
\usepackage{epsfig}
\usepackage{amssymb}
\usepackage{amsfonts}
\usepackage{bm}

\newcommand\bz{\bar{z}}
\newcommand\mutotal{\mu_{\rm t}}
\newcommand{\arsinh}{\mathop{\mathrm{arsinh}}}
\newcommand\ii{\ensuremath{\rm i}}
\def\wisk#1{\ifmmode{#1}\else{$#1$}\fi}

\def\lsim {\wisk{_<\atop^{\sim}}}

\begin{document}
\include{journaldef}

\title{Fast Computational Convolution Methods For Extended Source
Effects In Microlensing Lightcurves.}

\author{Hans J. Witt\altaffilmark{1}, F. Atrio-Barandela\altaffilmark{2}}
\altaffiltext{1}{Im Hollergrund 76, 28357 Bremen, Germany; h\_witt@gmx.de}
\altaffiltext{2}{F{\'\i}sica Te\'orica, Universidad de Salamanca,
37008 Salamanca, Spain; atrio@usal.es}

\begin{abstract}
Extended source effects can be seen in gravitational lensing events when sources
cross critical lines. Those events probe the stellar intensity profile and could be
used to measure limb darkening coefficients to test stellar model predictions.
A data base of accurately measured stellar profiles
will be needed to correctly subtract the stellar flux in
planetary transient events. The amount of data that is being and will
be produced in current and future microlensing surveys, from the space
and the ground, requires algorithms that can quickly compute light curves
for different source-lens configurations. Based on the convolution method we
describe a general formalism to compute those curves for single lenses.
We develop approximations in terms of quadratures of elliptic integrals
that we integrate by solving the associated first order differential equations.
We construct analytic solutions for a limb darkening and, for the first time,
for a parabolic profile that are accurate at the $\sim 1-3\%$ and $0.5\%$
level, respectively. These solutions can be computed orders of magnitude faster than
other integration routines. They can be implemented in pipelines processing
large data sets to extract stellar parameters in real time.
\end{abstract}


\section{Introduction}

Limb darkening is the variation of the specific intensity of a star from its
center to its limb. The main probes of the effect in
distant stars have been the light curves of eclipsing-binary systems
(see \citealt{Zola15}). Optical interferometry has allowed direct
imaging of those few stars whose angular diameter could be resolved
\citep{Hestroffer97, Aufdenberg05}. Exoplanetary transits are potentially
very useful to determine stellar profiles although empirical values
derived from data show poor agreement with atmospheric models
\citep{Howarth11}. Equally relevant is that these
transits provide accurate measurements of exoplanet
properties. To study the atmosphere of transiting planets accurate
limb-darkening stellar models are needed to correctly subtract the stellar flux
\citep{Morello17}. Therefore, different and independent data of the
limb-darkening effect are required to correctly determine the properties
of planetary atmospheres and to test stellar models.
Microlensing has become a powerful observational tool with
different astrophysical applications since the observation of the
first event \citep{Mao12}. In particular,
gravitational lensing is sensitive to the effect of limb darkening \citep{Witt95}.
Extended source effects can be seen when the source crosses a caustic;
the parameters of the star intensity profile can be measured from these
events and can be compared with the predictions of stellar models.
They were first seen in high magnification single-lens events
\citep{Alcock97,Jiang04,Fouque10,Zub11}. These events, although rare,
are important as they provide a way to measure the lens mass \citep{An02,Jiang04}.
\cite{Albrow99} were the first to determine the limb-darkening
coefficients of a K giant star using microlensing, the first ever
for a star in the Galactic bulge. \cite{Afonso00} measured the limb-darkening
coefficients for a metal-poor A star, in five bands from I to V, in a
binary lens event. Sources crossing caustic lines in the lens plane are
more common in multiple lens systems and extended effects could be more easily seen
\citep{Afonso00,Albrow01,Choi12,Chung17,Han18}.
Satellite observations like Spitzer offer a different line of sight than ground
telescopes. In \cite{Chung17}, the center of the source passed very close to
the projected position of the lens as seen by Spitzer
and the size of the star could be measured from a single point, although
no extended source effects were seen from Earth. Like in transient events,
the profile coefficients of stars measured by gravitational lensing
are not always in agreement with theoretical expectations \citep{Fouque10}.
The understanding of radiative transfer models needs to be
improved to fit microlensing observations \citep{Cassan06}, demonstrating
the need for additional data.

The effect of the source extension on the magnification pattern
can be readily computed by convolving the source profile with the
gravitational amplification of a point source since the magnification
of different sources are additive for a given lens.
Early calculations were centered on constant profiles
\citep{Gould94, WM94}. Convolution with an extended source was usually applied
in the vicinity of a single caustic (e.g. \citealt{SW87}) to compute the shape
of the light curve. Deconvolution has been used to recover the profile of
the lensed source \citep{Grieger91}. Although convolution was also applied to
compute gravitational magnifications \citep{WMS95}, no account to the numerical
method was given. In the near future we can expect a significant increment in data
and fast methods to obtain light-curves will be needed. Facilities
like the future Wide Field Infrared Survey Telescope ({\it WFIRST}) satellite,
currently under construction, will devote a significant fraction of its observing
time to the Galactic bulge, with a detection of
$\sim 50,000$ microlensing events being expected \citep{Spergel15,Gaudi19}.
The ESA {\it Euclid} satellite is also capable of high cadence observations on
a wide field of view, and in two colors, and it will also be a useful instrument
for microlensing. Being at a different L2 location than
{\it WFIRST}, events that are simultaneously observed by both satellites, and
possibly also from the ground, will be measured from different lines of sight,
facilitating the detection of extended source effects and allowing the lens mass to
be estimated directly from the data \citep{Penny19a}. The photometric error
of the {\it WFIRST} W149 filter would be $\sim 10^{-2}$ mag per exposure for a
AB$\sim 21$mag star \citep{Penny19b}, comparable or better than observations
from the ground. Processing those forthcoming data sets would require algorithms
that can obtain light curves of lensing events with an accuracy of $\sim 0.01$
mag or better. This is most efficiently done using analytic expressions
instead of light curves by numerical methods.

In this article we show that convolution offers a novel technique to
compute very efficiently extended source effects for single point-mass
lenses. It is an exact and computationally less expensive method than
integration of the image contour described in \citet{WM94} and \citet{GG97}.
Our approximations are more general than those used in \cite{Yoo04}
and \cite{Chung17}, are faster than currently available methods and can
be easily implemented in data analysis pipelines for the forthcoming data sets.
Briefly, in \S\ref{sec:Microlensing} we summarize the basic formalism
and we present the convolution method used to derive
analytic solutions; we also introduce the models of stellar limb-darkening
that will be considered in the article. In \S\ref{sec:single} we present our
estimates for single lens events, in terms of simple quadratures,
that can be applied to any source profile. In \S\ref{sec:exact}
we compute the exact analytic expressions for a constant intensity profile
and in \S\ref{sec:approx} we derive approximations for
a parabolic and limb-darkening profiles that are accurate at the
level of $\le 0.005$mag and $\lsim 0.02-0.03$mag, respectively.
These analytic formulae can not be generalized to binary
or multiple lens system and an analysis of these cases
will be given elsewhere. Finally in \S\ref{sec:conclusions} we
summarize our results and present our conclusions.

\section{Basic Results and Source Profiles}
\label{sec:Microlensing}

A full account of the theory of lensing can be found in \cite{SEF92}.
The lens equation for point masses is more easily solved if expressed in terms
of complex quantities \citep{Wi90}.
Let $\zeta=\xi+\ii\eta$ denote the (complex) position of a point source
in the $(\xi,\eta)$-source plane, $z=x+{\ii}y$ its image in
the $(x,y)$-lens plane and $z_i=x_i+{\ii}y_i$ the positions of a
field of $n$ point mass stars of mass $m_i=M_i/M$ also in the lens plane.
For this configuration, the lens equation is given by
\begin{equation}
\zeta = z +\sum_{i=1}^n \frac{m_i}{\bz_i - \bz},
\label{eq:1}
\end{equation}
where $\bz$ is the complex conjugate of $z$.
The quantities $z$ and $\zeta$ are in units of
the Einstein radius of the deflector/lens plane and source plane,
respectively, and stellar masses $m_i$ can be expressed
in units of the total mass
$M$ so that $\sum_i m_i = 1$. The normalized (Einstein) units are given by
\begin{equation}
z_{\rm E} = \sqrt{4GM D_{ls} D_l \over c^2 D_s} \quad {\rm and} \quad
\zeta_{\rm E} = {D_s \over D_l} z_{\rm E} ,
\end{equation}
where $D_l$, $D_s$, $D_{ls}$ are the distances to the lens,
the source and the distance between the lens and the source,
respectively. The solutions of eq.~(\ref{eq:1})
are the image positions $z$ for each source position $\zeta$.

In this article we will express the stellar radii $r_s$
in units of the Einstein radius of the lens-source system.
A star in the bulge of physical radius $R$
will have a radius $r_s$ in the source plane
\begin{equation}
\frac{r_s}{\zeta_{\rm E}}\simeq 0.29\left(\frac{R}{400R_\odot}\right)
\left[\left(\frac{D_l}{5\rm Kpc}\right)\left(\frac{5\rm Kpc}{D_{ls}}\right)
\left(\frac{10\rm Kpc}{D_s}\right)\right]^{1/2} ,
\label{eq:rs_magnitude}
\end{equation}
with $R_\odot$ the solar radius.  The fiducial value $R\simeq 400R_\odot$ is
characteristic of C-rich giant stars \citep{Paladini11,vanBelle13} so stellar
radii of bulge stars will typically be in the range
$r_s\simeq 10^{-3}\zeta_{\rm E} - 0.3\zeta_{\rm E}$.

The effect of gravitational lensing on a background source is to magnify its intensity.
A field of $n$ point-mass lenses in the deflector plane will
produce $k$ images of a given point source located at position $\zeta$
in the source plane. The magnification of one image located at $z_j$ is given by
\begin{equation}
\mu_j = {1 \over \det J} = \left. \left( 1 - {\partial \zeta \over \partial \bz}
\overline{\partial \zeta \over \partial \bz} \right)^{-1} \right|_{z=z_j} .
\end{equation}
The total magnification of the background source is the sum of the
absolute magnification of each image,
\begin{equation}
\mutotal(\xi,\eta)=\sum_{j=1}^k\left|\mu_j(\xi,\eta)\right|
=\sum_{j=1}^k\left|\mu(x_j(\xi,\eta),y_j(\xi,\eta))\right| . \label{eq:4}
\end{equation}
Images and total magnification can be computed for each source location
$(\xi,\eta)$ so the relative motion of the lens and source will produce
a magnification pattern over the whole source plane area.

From eq.~(\ref{eq:4}) it follows that the magnification of an arbitrary number
of sources by the same set of lenses is additive. Once the magnification of a point-like
star, $\mu(\xi,\eta)$, is known the magnification pattern for an extended
star can be derived by convolving the point-like pattern with the extended
source profile. If the center of a background star of radius $r_s$ is located
at $(\xi_0,\eta_0)$ then its magnification is
\begin{equation}
  \mu_{\rm ext} (\xi_0,\eta_0) = \int_{-\infty}^{\infty} \int_{-\infty}^{\infty}
  \mu(\xi,\eta) s(\xi_0-\xi,\eta_0-\eta) d\xi d\eta =
    \int_{-r_s}^{r_s} \int_{-r_s}^{r_s}
  \mu(\xi-\xi_0,\eta-\eta_0) s(\xi,\eta) d\xi d\eta , \label{eq:conv}
\end{equation}
where $s(\xi,\eta)$ is the source profile.
If the source is not lensed, then
the magnification pattern will be $\mu(\xi,\eta)=1$ everywhere
and the magnification will remain unchanged for any source profile. Then,
it follows the normalization condition
\begin{equation}
V_s=\int_{-\infty}^{\infty}\int_{-\infty}^{\infty}s(\xi,\eta)d\xi d\eta=1 ,
\label{eq:norm}
\end{equation}
valid for any profile $s$.

\subsection{The Conservation of Magnification}

Fubini theorem guarantees that the magnification is conserved,
irrespectively of the source profile. The theorem states that the volume
enclosed by the convolved function is equal to the product of the single volume
enclosed by the two functions provided the integral exists. Although
the volume enclosed by the magnification is infinite, the volume enclosed by
$\mu(\xi,\eta)-1$ is finite. The integral in eq.(\ref{eq:conv}) remains valid
if we replace $\mu(\xi,\eta)$ by $\mu(\xi,\eta)-1$ since the total
magnification is $\mu(\xi,\eta) \geq 1$ everywhere (cf. \citealt{Schn84}).
We can write
\begin{equation}
V_{\mu > 1}=\int_{-\infty}^{\infty}\int_{-\infty}^{\infty}(\mu_{\rm ext}(\xi,\eta)-1)d\xi d\eta=
V_s\int_{-\infty}^{\infty}\int_{-\infty}^{\infty}(\mu_{\rm point}(\xi,\eta)-1) d\xi d\eta ,
\label{eq:7}
\end{equation}
where the identity follows from Fubini theorem and is valid for a field of
$n$ point-like lenses. From the normalization condition given in eq.~(\ref{eq:norm})
it follows that the enclosed volume above $\mu >1$ does not change for arbitrary source
profiles and is equal to that of a point-like source, and then, finite.
Then eq.~(\ref{eq:7}) provides a self-consistency check to verify the accuracy of the
estimated amplification $\mu_{\rm ext}$ for any source profile.

We can solve eq.~(\ref{eq:7}) for simple cases. Let $r_0=(\xi_0^2+\eta_0^2)^{1/2}$
denote the separation of the center of source to the projected position of the lens,
in the source plane. For a single point mass lens, the magnification is \citep{Ref64}
\begin{equation}
\mu(r_0)= \frac{1}{2}\left(\frac{r_0}{\sqrt{r_0^2+4}}+\frac{\sqrt{r_0^2+4}}{r_0}\right) .
\label{eq:singlemu}
\end{equation}
Introducing polar coordinates in eq.~(\ref{eq:7}) by defining $r=(\xi^2+\eta^2)^{1/2}$
we obtain
\begin{equation}
V_{\mu>1}=2\pi\int_0^{\infty} (\mu(r)-1)rdr =
\pi\left[r\sqrt{4+r^2}-r^2\right]_0^{\infty}=2\pi
\end{equation}
For more complicated lens configurations, eq.~(\ref{eq:7}) can not be solved analytically.
For instance, the light curve of a close binary is very different from that of single
lens located at the same distance. The differences are largest when the distance
from the center of the source to the lens, $r_0$, is of the same order of magnitude
or smaller than the projected separation of the binary lens $d_{\rm lens}$.
On the contrary, when $r_0\gg d_{\rm lens}$ the magnification pattern is
that of a single lens. We solved eq.~(\ref{eq:7}) for several binary configurations
and carry out the integration from the center of mass of the binary to a distance $r_{\max}$
ten times the maximum of the binary lens separation and the radius of the source
star: $r_{\max}=10\,\max(r_s,d_{\rm lens})$. Our numerical estimates showed that,
as expected, the enclosed volume of the magnification $V_{\mu >1}$ remains constant
and depends only on the total mass of the system. We obtained
\begin{equation}  \label{eq:magvolume}
V_{\mu > 1}=2\pi(m_1+m_2)
\end{equation}
where $m_1$ and $m_2$ are the masses of the binaries in arbitrary units.
If we express the mass of each component in units of the total
mass of the system, then $V_{\mu > 1}=2\pi$ again. We conjecture that for
a general lensed point mass system the volume of the magnification pattern will
always depend on the total mass of the system $V_{\mu > 1}=2\pi \sum_i m_i$.

\subsection{Signed Magnification}

A direct consequence of the convolution integral for lens models (eq.~\ref{eq:conv})
is that in certain areas where the sum of signed magnification for point
sources is constant the relation is the same for extended sources.
Let us assume that we have a caustic network with an area
\begin{equation} \label{eq:signedmag}
c_{\mu}=\sum_i\mu_i^{+}-\sum_j\mu_j^{-}={\rm const.}
\end{equation}
where $\mu_i^{+}$ is the magnification of the image of positive parity
and $\mu_i^{-}$ the magnification of the image of negative parity.
By applying eq.(\ref{eq:conv}), for an extended source we can write
\begin{equation}
\sum_i\mu_{i, {\rm ext}}^{+} - \sum_j \mu_{j, {\rm ext}}^{-}
=\int_{-\infty}^{\infty} \int_{-\infty}^{\infty}c_{\mu} s(\xi,\eta)d\xi d\eta
= c_{\mu}
\end{equation}
since we may exchange the integral and sum of each magnification.
In particular for a single point mass lens we have
$c_{\mu}=1$ everywhere and for a binary lens we have $c_{\mu}=1$ inside
the caustic(s) (see \citealt{WM95}, \citealt{WM00}, \citealt{HE01}).
Let $A_i^{+}$ be the area of the lensed image of positive parity and
$A_j^{-}$ the area of the lensed image of negative parity. For the single
point mass and for the binary lens if the area of the source $A_s$ is located
inside the caustic, it follows that
\begin{equation}
\sum_i A_i^{+} - \sum_j A_j^{-} = A_s.
\end{equation}
When computing light curves for extended sources by a contour method
(cf. \citealt{GG97}), this equation provides a useful check on the accuracy
of the numerical integration over the extended sources. Alternatively, it
can be used to reduce the integration on the images of positive parity since
this would be sufficient to obtain the total magnification.

\subsection{Source Profiles}\label{sec:profile}

The intensity of a star can be very well modeled with a general
limb-darkening profile of the form (e.g. \citealt{Al73}, \citealt{CDG95})
\begin{equation}
\frac{I(r)}{I(0)}=1-u_1-u_2+u_1\sqrt{1-\frac{r^2}{r_s^2}}
+u_2\left(1-\frac{r^2}{r_s^2}\right) .
\label{eq:int_profile}
\end{equation}
The coefficients $u_1$ and $u_2$ depend on the observed
(wavelength) band of the star and may differ quite strongly
(for instance, see \citealt{Fouque10}).  Those quantities
are not independent; assuming that the intensity profile
(eq.~\ref{eq:int_profile}) is everywhere positive and monotonically
decreasing from the center to the limb, these coefficients
verify $u_1>0$, $u_1+u_2<1$ and  $2u_2+u_1>0$ \citep{Kipping13}.
The source profile normalized according to eq.~(\ref{eq:norm}) is
\begin{equation} \label{eq:starprofile}
s_{\rm star}(r)=
\frac{6(1-u_1-u_2)s_{\rm disk}(r)+4u_1s_{\rm limb}(r)+3u_2s_{\rm para}(r)
}{6-2u_1-3u_2} ,
\end{equation}
with
\begin{eqnarray}
s_{\rm disk}(r)&=&\frac{1}{\pi r_s^2} , \label{eq:starprofiles1}\\
s_{\rm limb}(r)&=&\frac{3}{2\pi r_s^2}\sqrt{1-\frac{r^2}{r_s^2}} , \label{eq:starprofiles2}\\
s_{\rm para}(r)&=&\frac{2}{\pi r_s^2}(1-\frac{r^2}{r_s^2}) ,\label{eq:starprofiles3}
\end{eqnarray}
where each of these three profiles obeys individually the normalization condition of
eq.~(\ref{eq:norm}). In the following sections
we will use these profiles to illustrate our analysis.

\section{The single point mass lens} \label{sec:single}

For a single point mass lens, the magnification depends only on the distance
between the source and the projected position of the lens (eq.~\ref{eq:singlemu}).
If the latter is at the origin of coordinates and $\bm{r}_0=(\xi_0,\eta_0)$,
$\bm{r}=(\xi,\eta)$ denote
the coordinates of the center of the source and of an arbitrary point on its surface,
respectively, then eq.~(\ref{eq:conv}) can be written in polar coordinates as
\begin{equation} \label{eq:integral}
\mu_{\rm ext}(r_s,r_0)=\int_{0}^{2\pi}\int_0^{r_s}s(r,\varphi)
\mu(|{\bm{r}-\bm{r}_0}|) r dr d\varphi ,
\end{equation}
being $|\bm{r}-\bm{r}_0|=(r^2+r_0^2-2rr_0\cos\varphi)^{1/2}$ and
$r_s$ the radius of the star. Computing the double integral
on a grid gives profiles with rigging whose amplitud
scales as $N_p^{-1/2}$, being $N_p$ the number of points
on the grid. The rigging affects only those trajectories
where the lens moves across the surface of the star and
to improve the accuracy requires to increase the number of points,
increasing the computational time.

Smoother light curves are obtained when
the stellar intensity profiles are spherically symmetric. In this case,
(see Appendix~\ref{sec:integral} and also \citealt{Witt95},
\citealt{Heyrovsky03}) the angular part can be integrated to obtain the
amplification in terms of a single quadrature of elliptical functions
\begin{equation} \label{eq:exint}
\mu_{\rm ext}(r_s,r_0)=\int_0^{r_s}
k\sqrt{\frac{r}{r_0}}[(r-r_0)^2\Pi(n,k)+2K(k)]s(r)dr ,
\end{equation}
with
\begin{equation} \label{eq:nk}
n=\frac{4r_0r}{(r_0+r)^2}, \quad k=\sqrt{\frac{4n}{4+(r_0-r)^2}} ,
\end{equation}
In eq.~(\ref{eq:exint}) $K(k)$, $\Pi(n,k)$ denote the complete elliptic integral
of the first and third kind.  The integral has a singularity at $k=1$
or $r=r_0$ so that care needs to be taken to obtain accurate results.
Gauss-Legendre integration over $10^3$ points introduces rigging at the
2-3\% level. We used the \citet{Press02} routines for improper integrals
{\it qtrap.f } and {\it midpnt.f }
for more accurate results. To reduce the amplitude of the rigging below
the 0.1\% level we run subdivisions up to JMAX=20 with an accuracy of
EPS=$10^{-5}$. This high precision slows down the code compared with
other integration routines.

\begin{figure}
\centering\epsfxsize=\textwidth \epsfbox{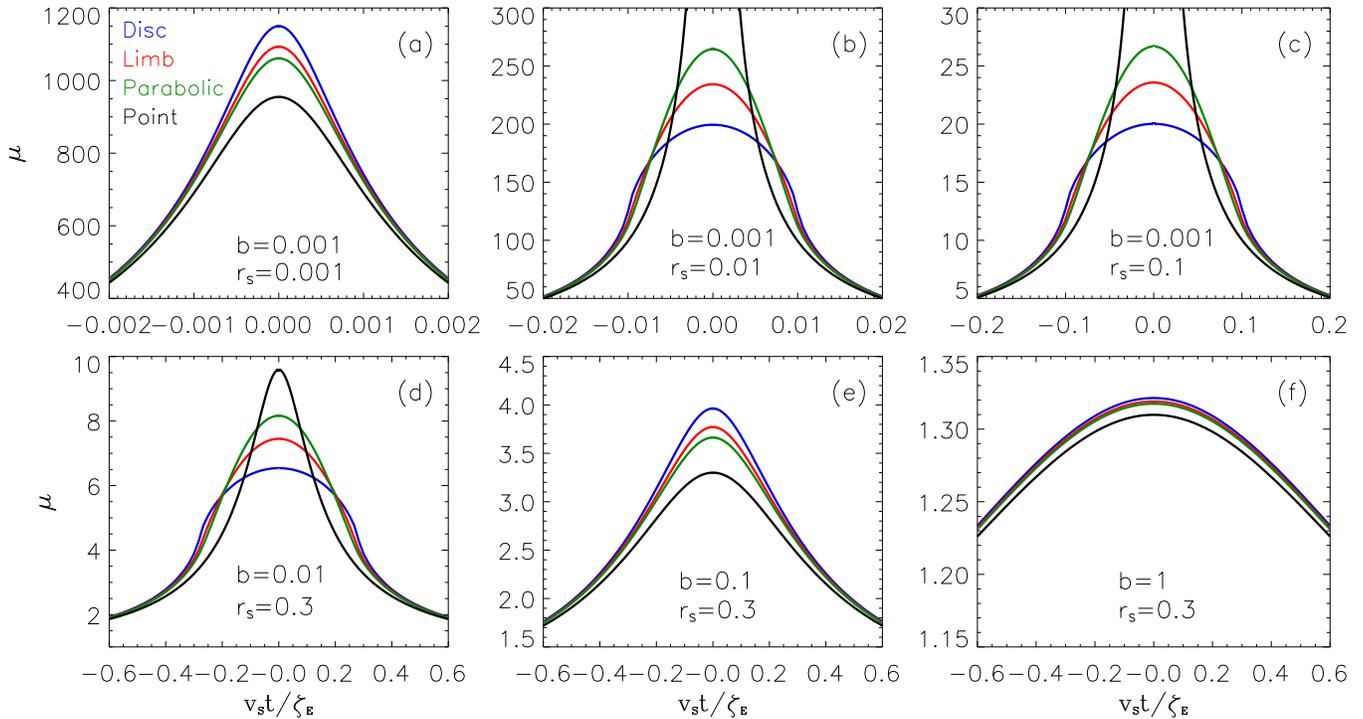}\hfil
\vspace*{-2cm}
\caption{\small
Gravitational lensing magnification of a point-like (solid black line)
and an extended source with intensity constant (blue), limb-darkening (red)
and parabolic (green) profiles. The different panels correspond to different
impact parameters $b$ and different stellar radii $r_s$, as indicated.
Distances and radii are given in units of the Einstein radius in the source
plane.
}
\label{fig:fig1}
\end{figure}

In Fig.~\ref{fig:fig1} we represent the magnification of a point source
and of the three extended source profiles given in
eqs.~(\ref{eq:starprofiles1}-\ref{eq:starprofiles3}), computed with
routines for improper integrals.
Depending on the impact parameter and stellar radius the magnification could
be larger or smaller than that of a point source. In all cases, the
parabolic profile was the closest to the point source and the constant profile
was the furthest. For instance, in plots (a,e,f) curves at the peak correspond,
from top to bottom, to the amplification of
a point source (black line) and extended sources with constant (blue),
limb-darkening (red) and parabolic (green) profiles. In (b,c,d), the color
code is the same but the ordering of the magnification curves is the reverse.
In the x-axis, $v_s$ is the relative velocity of the source
and the lens, and $t$ is the time of observation. The distance $r_0$ from
the center of the source to the projected position of the lens is given by
$r_0=[(v_st)^2+b^2]^{1/2}$, being $b$ the impact parameter. Distances and stellar
radii are given in units of the Einstein radius on the source plane, $\zeta_{\rm E}$.
Plots (a-c) correspond an impact parameter $b=10^{-3}$ and
stellar radii $r_s=10^{-3},10^{-2},0.1$; plots (d-f) correspond to impact parameters
$b=(0.01,0.1,1)$ and stellar radius $r_s=0.3$.
The largest differences between magnification of point and extended
sources with any of the three profiles occur when the lens is within
the projected surface of the star. Once the lens is outside
the magnification decreases, extended source effects are diluted
and the amplification coincides with that of a point source.

\section{Exact Analytic solutions.}
\label{sec:exact}

Faster numerical methods require the full integration of eq.~(\ref{eq:integral}).
Exact solutions for particular cases can be found by solving the differential
equation associated to eq.~(\ref{eq:exint})
\begin{equation}
\frac{d\mu(r,r_0)}{dr}=k\sqrt{\frac{r}{r_0}}[(r-r_0)^2\Pi(n,k)+2K(k)]s(r) ,
\label{eq:diff}
\end{equation}
In Appendix~\ref{sec:a_i} we show that this differential equation can be solved
exactly for the constant intensity profile $s(r)=s_{\rm disk}(r)=1/\pi r_s^2$.
The solution is given in terms of elliptic functions and has the form
\begin{equation}
\mu_{\rm disk}(r,r_0)=a_1(r)K(k)+a_2(r)E(k)+a_3(r)\Pi(n,k)
\label{eq:mudisk}
\end{equation}
with
\begin{equation}
a_1(r_s) = { k(r_s) \over \sqrt{r_0 r_s} } {(r_s^2-r_0^2)p(r_s) \over \pi r_s^2}
\, , \quad
a_2(r_s) = { \sqrt{r_0 r_s} \over k(r_s)} {q(r_s) \over \pi r_s^2}
\quad {\rm and} \quad
a_3(r_s) = { k(r_s) \over \sqrt{r_0 r_s} } \frac{(r_0-r_s)^2 (r_s^2+1)}{2 \pi r_s^2}
\label{eq:a_coefficients}
\end{equation}
where
\begin{equation}
p(r)= \frac{1}{8} (8+r_0^2-r^2) \quad {\rm and} \quad q(r) = 2 .
\label{eq:pq}
\end{equation}
The coefficients of eqs.~(\ref{eq:a_coefficients}, \ref{eq:pq}) were first
derived by \citet{WM94}. Elliptic integrals can be evaluated very efficiently
(for instance, with the routines given in \citealt{Press02}) and this solution
can be computed very quickly since no integration is required.

Fig~\ref{fig:fig1} demonstrates that the gravitational magnification
of all profiles converge when the source is far from
the lens. Eventually, the parabolic and limb darkening profiles
match smoothly the disc profile given by
eqs.~(\ref{eq:mudisk}-\ref{eq:pq}). We will assume that a general solution
of eq.~(\ref{eq:diff}) can be represented by the same
functional expression but with different coefficients or possibly functions
$(a_1,a_2,a_3)$, to be determined. Introducing this ansatz in eq.~(\ref{eq:mudisk})
gives a set of three coupled differential equations for the coefficients
$a_i(r)$ that needs to be solved for each specific profile.
To fix the constants of integration, the solution has to converge
to some limiting cases. For instance, the magnification
reduces to that of a point source when $r_s \rightarrow 0$
\begin{equation} \label{eq:exintrs0}
\lim_{r_s\rightarrow 0}\mu_{\rm ext}(r_s, r_0)=\frac{2+r_0^2}{r_0\sqrt{4+r_0^2}} .
\end{equation}
The differential equations for the coefficients given in Appendix~\ref{sec:a_i}
are rather complicated. For the limb and parabolic case
we found solutions only at two specific configurations,
when star and lens are perfectly aligned, $r_0=0$, and when the lens is at
the edge of the star, $r_0=r_s$. When $r_0=0$ eq.~(\ref{eq:exint}) becomes
\begin{equation} \label{eq:exint0}
\lim_{r_0\rightarrow 0}\mu_{\rm ext}(r_s,r_0)=
\frac{\pi}{2}(a_1(r_s)+a_2(r_s)+a_3(r_s)) .
\end{equation}
If $r_0=0$ or $r_0=r_s$ the elliptic integral simplify considerably (see
eqs.~\ref{eq:limEKPi}, \ref{eq:relPiE}) and we derived simple series expansions
that converge rather quickly when $r_s\ll 1$, the most common case.
These expressions are given in Appendix~\ref{sec:a_i}.

The results at these two locations, although limited, are still  informative.
Since the difference between profiles is largest when source and lens are
perfectly aligned, we can derive an upper bound on the contribution of
each profile to the overall magnification. Second, when
the lens is on the edge of the star extended source effects start
to be noticeable. If only a few (or just one) data point are available
it is useful to have an accurate estimate of the difference between
profiles at physically relevant locations.
The differences between two source profiles are given by
\begin{equation}
m_{\lambda_1}-m_{\lambda_2}=
-2.5\log\left(\frac{\mu_{\lambda_1}}{\mu_{\lambda_2}}\right) ,
\end{equation}
provided that they have the same intensity at wavelengths $\lambda_1$ and
$\lambda_2$. Otherwise, an additional factor needs to be included which
result in a constant magnitude between the same source at different wavelengths.
In the limit $r_0=0$, $\mu \propto r_s^{-1}$ and, by taking ratios, the stellar
radius cancels out.  The maximum possible magnitude difference between the
three extended profiles is independent of the radius.
From eqs.~(\ref{mudisk0}, \ref{mulimb0}, \ref{mupara0}) we have
\begin{eqnarray}
\label{eq:estimate1}
m_{\rm disk}-m_{\rm para} &=&
-2.5\log(\frac{\mu_{\rm disk}(r_s,0)}{\mu_{\rm para}(r_s,0)})\la 0.31\, {\rm mag}, \\
m_{\rm disk}-m_{\rm limb} &=&
-2.5\log(\frac{\mu_{\rm disk}(r_s,0)}{\mu_{\rm limb}(r_s,0)})\la 0.18\, {\rm mag} .
\nonumber
\end{eqnarray}
The magnitude difference with respect to a point source can not be
given since when perfectly aligned with the lens, the amplification diverges.

\begin{figure}
\centering\epsfxsize=\textwidth \epsfbox{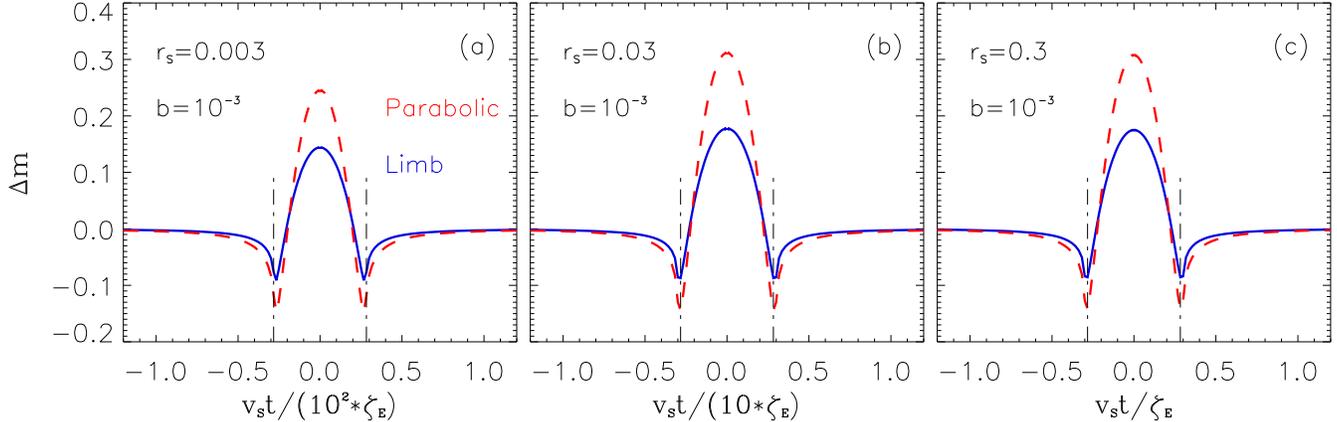}
\vspace*{-6.5cm}
\caption{\small Differences in magnitude of the limb (blue,
solid line) and parabolic (red, dashed line) profiles with respect
to the disk profile for different stellar radii.
Distances and stellar radii are measured in units of the Einstein radius
in the source plane, $\zeta_{\rm E}$. The impact
parameter is the same in the three panels. The vertical dot-dashed
lines indicate the position when the lens is at the edge of the star.
\vspace*{.5cm}
}
\label{fig:fig2}
\end{figure}

In Fig.~\ref{fig:fig2} we represent $m_{\rm disk}-m_{\rm limb}$ (solid
blue line) and $m_{\rm disk}-m_{\rm para}$ (dashed red line) for
different stellar radii. The vertical dot-dashed lines correspond to
$r_0=r_s$. When $b\approx 0$ and $r_s\gg b$, then $\Delta m$ are very
close to the limits given in eqs.~(\ref{eq:estimate1}).
The magnitude difference decreases with increasing distance and reaches
a minimum at $r_0=r_s$. The series expansion given in
Appendix~\ref{sec:a_i} can be used to estimate the differences
between profiles. From eqs.~(\ref{eq:b_rs_disc},
\ref{eq:b_rs_limb},\ref{eq:b_rs_para}) we have
\begin{eqnarray}
\label{eq:estimate2}
m_{\rm disk}-m_{\rm point}&=&
-2.5\log(\frac{\mu_{\rm disk}(r_s,r_s)}{\mu_{\rm para}(r_s,r_s)})
\approx 0.26\, {\rm mag}, \nonumber \\
m_{\rm disk}-m_{\rm para} &=&
-2.5\log(\frac{\mu_{\rm disk}(r_s,r_s)}{\mu_{\rm para}(r_s,r_s)})
\approx -0.13\, {\rm mag},\\
m_{\rm disk}-m_{\rm limb}&=&
-2.5\log(\frac{\mu_{\rm disk}(r_s,r_s)}{\mu_{\rm para}(r_s,r_s)})
\approx -0.08\, {\rm mag},
\nonumber
\end{eqnarray}
also in agreement with the results shown in Fig~\ref{fig:fig2}. Notice that at
a distance $r_0=r_s$ the magnification for the limb darkening and
parabolic profiles is smaller than that of a constant disk since for
the latter a larger fraction of the stellar flux is closer to the lens.

When the effect of an extended source is measured from very few or
just one single point (as in \citealt{Chung17}) one can not expect
to measure the relative contributions of different profiles with significant
accuracy. The estimates given in eqs.~(\ref{eq:estimate1}, \ref{eq:estimate2}),
together with the inequalities given in \S\ref{sec:profile} could still be used to
test if the stellar profile deviates from a constant even though the data could
not determine the coefficients $u_1$ and $u_2$ of eq.~(\ref{eq:starprofile}).

\section{Approximate analytic solutions.}
\label{sec:approx}

The differential equations obtained by the technique described in section
\S\ref{sec:exact} and Appendix \ref{sec:a_i} are very difficult to solve
for a generic profile. Approximate analytic solutions can be obtained
by a similar, albeit more simplified, technique. Fig.~\ref{fig:fig1} demonstrates that
when the distance from the center of a source to the projected position of a lens
is larger than a few stellar radii all light curves match the constant stellar profile.
Since the disk solution is analytic, we are only required to construct approximate
and accurate solutions in the range where the magnifications deviate,
approximately $r_0\le 3r_s$. When the stellar profiles depend only on the radial
distance $r$, we can compute the amplification of extended sources
by Taylor expanding around the origin the magnification given in eq.~(\ref{eq:singlemu})
\begin{equation}
\mu=\frac{2+r^2}{r\sqrt{4+r^2}} = \sum_{n=0}^{\infty} c_n r^{2n-1} \approx
 \frac{1}{r} + \frac{3}{8} r - \frac{5}{128} r^3 + \frac{7}{1024}r^5-\cdots
\label{eq:series}
\end{equation}
This expansion converges for $|r|< 2$ (in units of the Einstein radius) and
can not be applied to larger distances. The integration of
eq.~(\ref{eq:integral}) can be carried out term by term giving
\begin{equation}
\mu_{\rm ext} (r_s, r_0) = \sum_{n=0}^{\infty} c_{n} I_n (r_s, r_0) ,
\label{eq:mu_series}
\end{equation}
where $c_{n}$ are the coefficients in eq.~(\ref{eq:series}).
If we express the impact parameter in units of the source radius
$r_0 = u r_s$ and set the integration variable to $t=r/r_s$ then
\begin{equation}
I_n (r_s, u r_s) =  \frac{r_s^{2n-1}}{\pi} \hat{I}_n (u)
\end{equation}
with
\begin{equation}
\hat{I}_n (u) =  \int_{0}^{2\pi}
\int_{0}^{1}\frac{(u^2-2tu \cos\varphi+t^2)^{n}}{\sqrt{u^2-2tu \cos\varphi+t^2}}
\hat{s}(t) t dt d\varphi
\label{eq:In}
\end{equation}
and $\hat{s}(t) = s(r_s t) \pi r_s^2$. In Appendix~\ref{sec:series_expanion}
we demonstrate that the integrations in eq.~(\ref{eq:In}) can be expressed in
terms of elliptic integrals and pairs of functions $e_n(t), f_n(t)$ and take the form
\begin{equation}
\hat{I}_n(u)= (1-u) f_n(u)K\left(\frac{2\sqrt{u}}{1+u}\right)
+(1+u) e_n(u)E\left(\frac{2\sqrt{u}}{1+u}\right) .
\label{eq:I_ngeneral}
\end{equation}
Our treatment provides analytical approximations for the disk, parabolic and
limb profiles in terms of elliptic integrals of the first and second kind.
The solutions given in Appendix~\ref{sec:series_expanion} have to be
evaluated at $t=1$, at the radius of the star. The dominant contribution is
$n=0$ and higher orders contribute more with increasing source mass and
distance to the lens. For completeness we present the results up to $n=2$.
\begin{table}[t!]
\begin{tabular}{|l|r|r|r|r|r|r|}
\hline
      & $f_0$ & $e_0$ & $f_1$ & $e_1$ & $f_2$ & $e_2$ \\
\hline
$a_0$ & 692583091200 & 1385166182400 & 400313026713600 & 400313026713600 & 144513002643609600 &
144513002643609600 \\
$a_1$ & 1339448033263& 3894814720117 & -104737154144542& 586396203521047 & -41601097496530466 &
129881415373196381\\
$a_2$ & 312948159897 & 243184124641  & 118605243978153 & 130620128454965 & 25747524261728378 &
302061597305296826\\
$a_3$ & 72606018032  & 41938764464   & -12149180921700 & 2732363448097   & 16403635906248033 &
17159057255113677\\
$a_4$ & 502361257    & 252468377     & 19577824287     & 233199164080    & -760401005566596 &
-717449096541348 \\
$a_5$ & 4109895      & 1997995       & 5187655332      & 840367513       & 683702230503 &
569981448915 \\
$a_6$ & 79511        & 41279         & 7438983         & 4461547         & 29226547953 &
87184849284\\
$a_7$ & 835          & 527           & 111191          & 66815   	 & 120111823 &
82551251\\
$a_8$ & 308          & 308           & 963	       & 655		 & 1392911 & 931319\\
$a_9$ & 	     &		     & 308	       & 308 		 & 1107 & 799\\
$a_{10}$&	     &		     &		       &	         & 308 & 308 \\
\hline
\end{tabular}
\label{table:tab1}
\caption{Coefficients of the series expansion of the limb darkening profile out to 7th order.
\vspace*{.5cm}}
\end{table}

\begin{itemize}
\item{} For the disk profile $\hat{s}_{\rm disk}(t)=1$ we have
\begin{eqnarray}
f_{\rm 0, disk}(u) &=& 2 ,\hspace*{97pt}
e_{\rm 0, disk}(u) = 2 ,\nonumber \\
f_{\rm 1, disk}(u) &=& -\frac{2}{9} (1-u^2) , \hspace*{60pt}
e_{\rm 1, disk}(u) = \frac{2}{9} (7+u^2) ,\label{eq:efdisk} \\
f_{\rm 2, disk}(u) &=& -\frac{2}{75} (1-u^2)(13+3u^2),  \quad
e_{\rm 2, disk}(u) = \frac{2}{75} (43+82 u^2+3u^4). \nonumber
\end{eqnarray}

\item{} For the parabolic profile $\hat{s}_{\rm para}(t)=2(1-t^2)$ we obtain
\begin{eqnarray}
\label{eq:efparabolic}
f_{\rm 0, para}(u) &=& \frac{16}{9}(1 - u^2) , \hspace*{98pt}
e_{\rm 0, para}(u) = \frac{16}{9}(2-u^2), \nonumber \\
f_{\rm 1, para}(u) &=& -\frac{16}{225} (1-u^2)(4-u^2),  \hspace*{58pt}
e_{\rm 1, para}(u) = \frac{16}{225} (19+6u^2-u^4), \\
f_{\rm 2, para}(u) &=& -\frac{16}{3675} (1-u^2)(53+30u^2-3u^4),  \quad
e_{\rm 2, para}(u) = \frac{16}{3675} (158+449 u^2+36u^4-3u^6). \nonumber
\end{eqnarray}

\item{} For the limb-darkening profile $\hat{s}_{\rm limb}(t)=
(3/2)(1-t^2)^{1/2}$ the solution can not be computed in closed form.
We also need to expand the profile in Taylor series and find a
solution at each order. The technique quickly produces cumbersome
expressions. Increasing the order in the Taylor expansion improves the
accuracy at $r_0\simeq b$, but the resulting series diverge faster.
We found that the results up to the 7th order expansion provided
the most accurate approximations in the range $r_0=(0,2.5r_s)$
after which the solution can be matched to the exact disk profile
with minimal error. In this case,
\begin{eqnarray}
\label{eq:coeff_limb}
\nonumber
f_{\rm 0, limb}(u) &=& (a_1-2u^2(a_2+ a_3u^2 + 64 u^4 (a_4+ 64 u^2 (
a_5+ 28 u^2 (a_6+ 48 u^2 (a_7+ a_8 u^2))))))/a_0, \\
\nonumber
e_{\rm 0, limb}(u) &=& (a_1- 4 u^2 (a_2+ a_3u^2 + 64 u^4 (a_4+
        64 u^2 (a_5+ 28 u^2 (a_6+ 48 u^2 (a_7+ a_8 u^2))))))/a_0, \\
f_{\rm 1, limb}(u) &=& (a_1+ a_2u^2 + a_3u^4 -
  64 u^6 (a_4+ a_5u^2 + 256 u^4 (a_6+ 28 u^2 (a_7+ 48 u^2 (a_8+ a_9u^2)))))/a_0,\\
\nonumber
e_{\rm 1, limb}(u) &=& (a_1+ a_2u^2 - 4 u^4 (a_3+ a_4u^2 +
     64 u^4 (a_5+ 64 u^2 (a_6+ 28 u^2 (a_7+ 48 u^2 (a_8+ a_9u^2)))))) /a_0,\\
\nonumber
f_{\rm 2, limb}(u) &=& (a_1+ a_2u^2 + a_3u^4 - a_4u^6 - 64 u^8 (a_5+
     4 u^2 (a_6+ 64 u^2 (a_7+ 28 u^2 (a_8 + 432 u^2 (a_9+ a_{10}u^2))))))/a_0,\\
\nonumber
e_{\rm 2, limb}(u) &=& (a_1+ a_2u^2 + a_3u^4 +a_4u^6 - 64 u^8 (a_5+ a_6u^2 +
     256 u^4 (a_7+ 28 u^2 (a_8+ 432 u^2 (a_9+ a_{10}u^2)))))/a_0 .
\end{eqnarray}
The coefficients $(a_0,\dots,a_{10})$ are given in Table~1.
A routine to compute these functions is available upon request.
\end{itemize}

\begin{figure}
\centering\epsfxsize=\textwidth \epsfbox{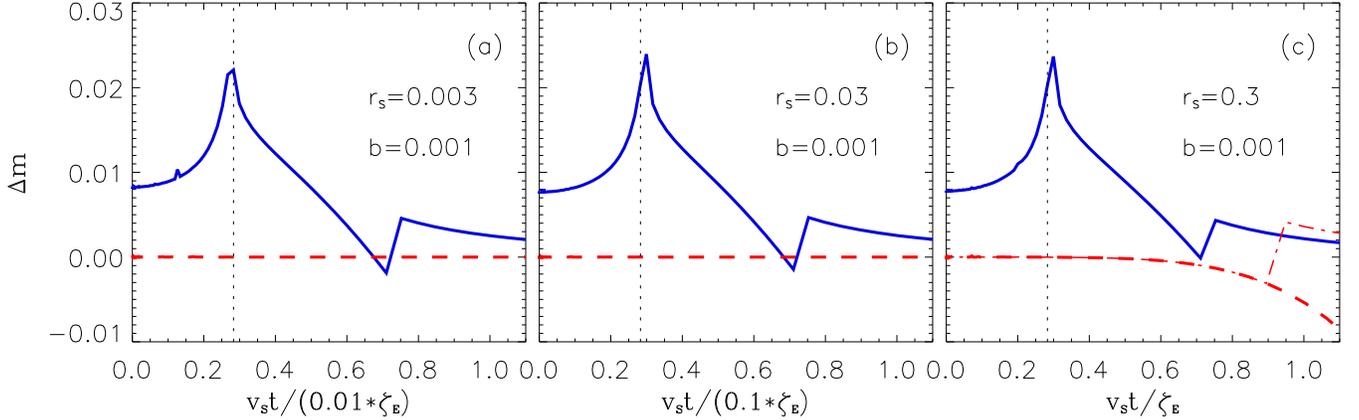}
\vspace*{-6.5cm}
\caption{\small
The blue solid and red dashed lines represent the magnitude difference between
the limb and parabolic profiles, respectively, computed from the approximations
of eqs.~(\ref{eq:coeff_limb}) and eqs.~(\ref{eq:efparabolic}) and from the
numerical integration of the improper integral of eq.~(\ref{eq:exint}), the former
extended out to $r_0=2.5r_s$ and continued with the disc solution given by
eqs.~(\ref{eq:mudisk}-\ref{eq:pq}).
The vertical dashed lines indicate the radius of the star. Panels
correspond to different stellar radii, as indicated.
In (c) the thin dot-dashed red line corresponds to a parabolic profile extended
out to $r_0=3r_s$ and continued with the exact disc solution as with the limb profile.
\vspace*{.5cm}
}
\label{fig:fig3}
\end{figure}

In Fig.~\ref{fig:fig3} we plot the magnitude difference $\Delta m=m_{\rm numerical}-
m_{\rm analytic}$ between the analytic approximations of eqs.~(\ref{eq:coeff_limb}) and
eqs.~(\ref{eq:efparabolic}) with respect to the numerical
integration of eq.~(\ref{eq:exint}). The dashed (red) and solid (blue)
lines correspond to limb and parabolic profiles. The limb analytic profile
is computed out to $r_0=2.5r_s$ and beyond that distance it is matched to
the exact constant disc expression of eqs.~(\ref{eq:mudisk}-\ref{eq:pq}).
In (a) the small wiggle of amplitude $\sim 10^{-3}$mag is due to
instabilities on the numerical solution. Stellar radii and the impact
parameter are given in units of the Einstein radius.

The zero order term in the Taylor expansion is always the dominant. For
sources with radii $r_s\le 0.01$, the first order contributes with 0.1\% to
the total magnification and the second order is negligible. For a star of
$r_s=0.3$, the first and second order term contribution is about 30\% and a 3\%
at large distances. The differences between the exact and approximated
parabolic profiles (dashed red line) is always negligible and only
for $r_s=0.3$ and at distances $v_st\ge 4r_s$ the error approaches $0.01$mag.
The amplitude is smaller than the numerical result indicating that terms
higher than the 2nd order in the expansion of eq.~(\ref{eq:mu_series})
are important. However, at those distances the exact disk
solution is very accurate and, like in the limb darkening case, is
simpler to switch to it instead of computing terms like $I_3$ or higher.
In Fig.~\ref{fig:fig3}c the thin dot-dashed red line shows the analytic solution
matched with the exact disc solution. We chose to match both solutions at
$r_0=3r_s$, which reduces the error in magnitude between the approximate
and exact solutions to less than $5\times 10^{-3}$mag.

The largest magnitude difference in the limb darkening case occurs at
$r_0\approx r_s$. The error was always smaller than $0.02$ mag except in
the narrow range $(0.27,0.30)$ where it was less than $0.03$ mag. Comparing with
the differences between profiles presented in Fig~\ref{fig:fig2} those
errors are 5-30\% of the difference between disk, parabolic and limb profiles,
proving that our approximations are sensitive to the different contributions.
In the case of the limb profile, we overestimate the correct
amplification and, consequently, $u_1$ in eq.~(\ref{eq:int_profile}) will
be underestimated. Once the range of parameter space that fits a light curve
has been identified, the bias can be corrected by fitting
numerically integrated profiles.

To summarize, our approximations give the effect of an extended source
in a microlensing event with errors $<3\%$ in the range $r_s\lsim 0.3$ and
are free from instabilities that affect numerical integration.
Computing light curves by solving eq.~(\ref{eq:integral}) on a
circular grid of $\sim 7,800$ points, eq.~(\ref{eq:exint}) using the
Gauss-Legendre on a linear grid of $10^3$ points, the same integral
using the {\it qtrap.f} and {\it midpnt.f} routines was $1.1\times 10^{3}$,
$430$ and $11\times 10^{3}$ times slower than using the analytic approximations
presented here. Our results generalize those of \cite{Yoo04} and the technique
can be extended to other profiles with even powers of the stellar radius.

\section{Conclusions}
\label{sec:conclusions}

We have developed an analytical formalism to compute the
light curves of extended sources gravitationally amplified by single lenses.
For the limb darkening profile we have obtained more accurate
analytic approximations than those currently being used in the literature.
We present even more accurate approximations for a parabolic profile.
To our knowledge, these are the first approximations in the literature
valid for this profile. To generate our approximations we first showed that
eq.~(\ref{eq:integral}) can be reduced to a single quadrature involving
elliptic integrals (eq.~\ref{eq:exint}). We integrated this expression
exactly for a constant stellar intensity profile by solving its
associated first order differential equation, given by eq.~(\ref{eq:diff}),
using the ansatz of eq.~(\ref{eq:mudisk}). Applying this formalism to
the limb-darkening and parabolic profiles we obtained accurate expressions
for two specific configurations: when the lens and source
were perfectly aligned and when the lens was on the boundary of the source.
These solutions were expressed in terms of the stellar radius $r_s$
and they converge very quickly for of $r_s\ll r_s$.

Next, we constructed approximations by Taylor expanding the magnification out to
second order. These approximations were valid for lens-source distances
$r_0\sim 3 r_s$. For the limb darkening profile, our approximations were
accurate to the 1\% level, except for a small interval when the lens is at
the edge of the stellar disk, that was only better than a 2-3\%. The net
effect is that the analytic solution will underestimate the $u_1$
coefficient in eq.~(\ref{eq:starprofile}). This shortcoming is not
very relevant since once extended source effects have been identified,
one can fit the data to light curves computed with more accurate numerical
methods but centered around the parameter space selected by the
analytic approximations. For the parabolic profile, this solution was
almost identical to the numerical integration
for a wide range of parameter space and only for very large
stars ($r_s\ge 0.3$) and at large distances ($r_0\ge 3.0$)
the error approached 1\% (see Fig.\ref{fig:fig3}).
For those configurations, the parabolic profile has already
converged to the constant disc solution and the error is negligible,
smaller than the expected photometric errors of forthcoming satellites.

In the next decade, {\it WFIRST}, {\it Euclid}
and telescopes from the ground will greatly increase the measured number
of microlensing events. Although their primary aim will be to search for
exoplanets and characterize their properties, they will also probe
stellar intensity profiles. Measuring limb darkening coefficients
at different wavelengths will help to improve the numerical models
of stellar atmospheres, a necessary step to study planetary atmospheres
in transient events and correctly subtract the stellar flux. The data analyses
will be helped by faster algorithms using analytically generated light curves.

\vspace*{1cm}
{\bf Acknowledgments}

We thank Shude Mao for comments on the manuscript.
F. A.-B. acknowledges financial support from Grants No.
PGC2018-096038-B-I00 (MINECO/FEDER) and No. SA083P17 from the Junta de Castilla
y Le\'on.

\appendix

\section{A. Angular integration of the magnification equation.}
\label{sec:integral}

In eq.~(\ref{eq:integral}) the integration of the angular variable $\varphi$ is
\begin{equation}
I =\int_0^{2\pi}\mu(|\bm{r}-\bm{r}_0|) d\varphi
  =2\int_0^{\pi} \frac{(2 + R^2 - 2 r r_0 \cos \varphi) d \varphi}{
\sqrt{R^2-2 r r_0\cos\varphi} \sqrt{4+R^2-2r r_0 \cos\varphi}} ,
\end{equation}
where $R^2=r^2+r_0^2$. The change of variables $t=\cos\varphi$ yields to
\begin{equation}
I = 2\int_{-1}^{1}\frac{\sqrt{b-t}}{\sqrt{a-t}\sqrt{1-t^2}} dt
  +\frac{2}{rr_0}\int_{-1}^1 \frac{dt}{\sqrt{(a-t)(b-t)}\sqrt{1-t^2}} ,
\end{equation}
where $a=(4+r^2+r_0^2)/2rr_0$ and $b=(r^2+r_0^2)/2rr_0$.
Using eqs.~(253.00,253.02) of \citet{BF71} we finally obtain
\begin{equation}
I=\frac{4(b-1)}{\sqrt{(a-1)(b+1)}}\Pi\left(\frac{2}{b+1}, k\right)
  +\frac{4 K(k)/(rr_0)}{\sqrt{(a-1)(b+1)}},
\end{equation}
with $k^2=[2(a-b)]/[(a-1)(b+1)]$. This expression gives
eq.~(\ref{eq:exint}).

\section{B. Differential Equations of the Coefficients}
\label{sec:a_i}

We will integrate eq.~(\ref{eq:exint}) by assuming that the solution
of the associated differential equation
(eq.~\ref{eq:diff}) has the functional form
\begin{equation}
\mu(r,r_0)=a_1(r)K(k)+a_2(r)E(k)+a_3(r)\Pi(n,k)
\label{eq:functional_expression}
\end{equation}
Using the properties of the complete elliptic functions given in
Appendix~\ref{sec:ellipticfct}, we obtain the following three
differential equations for the coefficients $a_i(r)$
\begin{eqnarray}
\frac{da_1}{dr}-(a_1+a_2)\frac{k^{\prime}}{k}&=&g_1(r) ,\label{eq:a1}\\
\frac{da_2}{dr}+a_2\frac{k^{\prime}}{k}+a_1\frac{k^{\prime}}{(1-k^2)k}&=&g_2(r) ,
\label{eq:a2}\\
\frac{da_3}{dr}+\left(\frac{1}{2r}+\frac{2}{(r_0-r)}
-\frac{k^{\prime}}{k}\right)a_3 &=& g_3(r) . \label{eq:a3}
\end{eqnarray}
The function $k=k(r)$ is defined in eq.~(\ref{eq:nk}), $k{\prime}=dk/dr$  and
\begin{eqnarray}
g_1(r) &=& \frac{a_3(r_0 +r)}{2r(r_0-r)}+2k\sqrt{\frac{r}{r_0}}s(r), \nonumber \\
g_2(r) &=&-\frac{(r_0+r)^2(4+r_0^2-r^2)a_3}{2r(r_0-r)^2(4+(r_0+r)^2)}, \\
g_3(r) &=& k\sqrt{\frac{r}{r_0}}(r_0-r)^2s(r) .\nonumber
\end{eqnarray}
Eq.~(\ref{eq:a3}) is not coupled to the other two equations and is the simplest to solve.
The solution of the homogenous equation, $a_{3,H}(r)$ can be found by integration
\begin{equation}
a_{3,H}=c_3\frac{(r_0-r)^2}{(r_0+r)\sqrt{4 +(r_0-r)^2}}=
c_3\frac{(r_0-r)^2k(r)}{4\sqrt{r_0r}} ,
\end{equation}
where $c_3$ is a constant. A solution of the inhomogeneous equation can
be found by variation of the constant. We write
\begin{equation}
a_3(r)=c_3(r)a_{3,H}=c_3(r)\frac{(r_0-r)^2 k(r)}{4\sqrt{r_0r}}
\end{equation}
By inserting this expression in eq.~(\ref{eq:a3})
we obtain $c_3^\prime(r) = 4 r s(r)$ whose solution is
\begin{equation}
c_3(r) = \int 4 r s(r) dr
\end{equation}
The solutions for the three different source profiles of
eqs.~(\ref{eq:starprofiles1}-\ref{eq:starprofiles3})
are presented below.

The coupled differential eqs.~(\ref{eq:a1},\ref{eq:a2}) are more difficult
to solve. We decoupled the two equations by differentiating again and the
solutions of the homogeneous equations are given in terms of elliptic integrals.
It is easy to verify that two independent solutions of the homogeneous
eqs.~(\ref{eq:a1}) and (\ref{eq:a2}) are: (1) $a_{11}=E(k(r))$, $a_{21}=-K(k(r))$
and (2) $a_{12}= (E(\hat{k})-K(\hat{k}))$, $a_{22}= K(\hat{k})$ with
$\hat{k} = \sqrt{1-k^2}$ (cf. \citealt{WW15}). The general homogenous solution is
\begin{eqnarray}
a_{1,H}(r)&=&c_1a_{11}+c_2a_{12}= c_1E(k)+c_2(E(\hat{k})-K(\hat{k})) \nonumber \\
a_{2,H}(r)&=&c_1a_{21}+c_2a_{22}=-c_1K(k)+c_2K(\hat{k}) ,
\end{eqnarray}
The solution of the inhomogeneous equation can be obtained by variation of
parameters $c_1$ and $c_2$. We write
\begin{eqnarray} \label{eq:a1a2res}
a_1(r) &=& c_1(r) a_{11} + c_2(r) a_{12} ,\nonumber \\
a_2(r) &=& c_1(r) a_{21} + c_2(r) a_{22} .
\end{eqnarray}
We insert this ansatz into eqs.~(\ref{eq:a1},\ref{eq:a2}) to obtain
\begin{eqnarray}
c_1^\prime a_{11} + c_2^\prime a_{12} &=& g_1 ,\nonumber \\
c_1^\prime a_{21} + c_2^\prime a_{22} &=& g_2 .
\end{eqnarray}
This equation can be readily solved for $c_1^\prime$ and $c_2^\prime$
and then integrated. Finally we can write
\begin{eqnarray} \label{eq:c1c2}
c_1(r) &=& \frac{2}{\pi} \int ( a_{22} g_1 - a_{12} g_2) dr \nonumber \\
c_2(r) &=& \frac{2}{\pi} \int ( -a_{21} g_1 + a_{11} g_2) dr .
\end{eqnarray}
since by the Legendre relation (see eq.(\ref{eq:Legendre})) we have
$a_{11}a_{22}-a_{12}a_{21}=\pi/2$. Then, from eq.~(\ref{eq:a1a2res}) we obtain
\begin{equation}
a_1(r) K(k(r)) + a_2(r) E(k(r))= c_2(r)\frac{\pi}{2}=\int [K(k(r))g_1(r)+E(k(r))g_2(r)]dr .
\end{equation}
Introducing this results into eq.~(\ref{eq:functional_expression}) we obtain
\begin{equation}
\int_0^{r_s} k \sqrt{\frac{r}{r_0}}[(r-r_0)^2\Pi(n,k)+2K(k)]s(r)dr=
a_3(r_s) \Pi(n,k) + \int [ K(k) g_1(r) + E(k) g_2(r) ] dr  .
\label{eq:a_3int}
\end{equation}
Since the integrals in eqs.~(\ref{eq:c1c2}) are rather involved
we will solve them by separating variables $r$ and $r_0$
in a series expansion. We define
\begin{equation} \label{eq:EKrel}
a_1(r)K(k(r)) + a_2(r)E(k(r))=
\frac{k(r)}{\sqrt{r_0 r}} \frac{(r^2-r_0^2)p(r)}{\pi r_s^2} K(k(r))
+\frac{\sqrt{r_0 r}}{k(r)} \frac{q(r)}{\pi r_s^2} E(k(r)) ,
\end{equation}
with
\begin{equation}
p(r)=\sum_{n=0}^{\infty} p_{2n}(r) r_0^{2n} \quad {\rm and} \quad
q(r)=\sum_{n=0}^{\infty}q_{2n}(r)r_0^{2n}\, .
\label{eq:pqn}
\end{equation}
By differentiating eq.~(\ref{eq:EKrel}) and after some algebra we
obtain two polynomial equations of the form
\begin{eqnarray}
32 r (r_0^2-r^2) p^\prime(r) -16(r_0^2+3r^2)p(r)+
(r_0^2-r^2)(4+r_0^2+3r^2)q(r) &=& h_1(r) , \nonumber \\
16  (4+r_0^2+3r^2) p(r)-(4+(r+r_0)^2)(4+(r-r_0)^2)[q(r)+2rq'(r)] &=& h_2(r) ,
\label{eq:polynomials}
\end{eqnarray}
with
\begin{eqnarray}
h_1(r) &=& [4 (r^2-r_0^2) c_3(r) - 64 r^2 s(r)] \pi r_s^2  ,\nonumber \\
h_2(r) &=& 4 (4+r_0^2-r^2) c_3(r) \pi r_s^2 \, .
\label{eq:h1h2}
\end{eqnarray}
Since the solution of the homogenous differential equation has the form
\begin{equation}
p_{H,0}(r) = c_1 \frac{\sqrt{4+r^2}}{r} - c_2\frac{\sqrt{4+r^2}}{r} \ln[r (4+r^2)]
\quad {\rm and} \quad
q_{H,0}(r) = \frac{-16 c_1}{r \sqrt{4+r^2}} +
16 c_2 \frac{(\ln[r (4+r^2)] +2)}{r\sqrt{4+r^2}}
\end{equation}
one obtains first an integral for $p_0(r)$ and $q_0(r)$ by variation of the
constants $c_1$ and $c_2$. The higher order $p_{2n}(r)$ and $q_{2n}(r)$
can be then obtained by recursive integration and by
using the homogenous solution.
Further if $h_1(r)$ and $h_2(r)$ have even power then the solutions $p(r)$ and
$q(r)$ of the inhomogeneous equations must have even power as well as
is the case of the three source profiles discussed here.

Solving the system of eqs.~(\ref{eq:a1}-\ref{eq:a3}) is rather complicated.
Only for a constant disk profile it is possible to obtain a simple exact
solution, given below. For other profiles, simple solutions exist for
specific configurations, when the approximations given in
eqs.~(\ref{eq:limEKPi},\ref{eq:relPiE}) are valid. The general (and cumbersome)
case can be solved in terms of power series expansions (eq.~\ref{eq:pqn})
but we shall not discuss it here. We defer to
Appendix~\ref{sec:series_expanion} the derivation of
accurate approximations for the limb and parabolic profiles.

\subsection{Disk of Constant Surface Brightness}

The profile of a disk of constant surface brightness is
simply $s(r)=1/(\pi r_s^2)$ and
\begin{equation}
c_3(r)= { (2 r^2+C_3) \over \pi r_s^2} ,
\end{equation}
where the constant $C_3$ needs to be determined by adequate limiting conditions.
For $a_3(r)$ we obtain
\begin{equation}
a_3(r_s)=
\frac{(2 r_s^2+C_3)}{\pi r_s^2}\frac{(r_0-r_s)^2 k(r_s)}{4\sqrt{r_0r_s}} .
\end{equation}
To fix the constant of integration, we take the
the limit $r_0\approx 0$. In this limit, the r.h.s of eq.~(\ref{eq:a_3int})) can
be transformed into a quadrature over elementary functions. When $r_s\ll 4$,
$k\approx \sqrt{n}\approx 0$ and we can apply the approximation given in
eq.~(\ref{eq:relPiE}). In the limit $r_0=0$ the
elliptic integrals are $K(0)=\Pi(0,0)=E(0)=\pi/2$ and
eq.~(\ref{eq:exint}) reduces to
\begin{equation}
\mu_{ext}(r_s,0)=2\pi\int_0^{r_s}\frac{(r^2+2)}{\sqrt{4+r^2}}s(r) dr .
\label{eq:mu_limit}
\end{equation}
For a constant surface brightness
\begin{equation} \label{mudisk0}
  \mu_{\rm disk}(r_s, 0) = {2 \over r_s^2} \int_0^{r_s}
  { (r^2+2) \over \sqrt{ 4+r^2 }} dr = {\sqrt{4+r_s^2} \over r_s }
  \approx \frac{2}{r_s}+ \frac{r_s}{4}- \frac{r_s^3}{64} +...
\end{equation}
Comparing this series with the integration of the l.h.s. of eq.~(\ref{eq:a_3int})
given in eqs.~(\ref{eq:EKrel}-\ref{eq:h1h2})
fixes the integration constant to the value $C_3=2$. Inserting the polynomials
$p(r)$ and $q(r)$ into
eqs.~(\ref{eq:polynomials}) triggers rather simple results for $p_0(r)$ and
$q_0(r)$. Further integration yields $p_2(r)=1/8$ and $q_2(r)=0$ and the
series terminates. The final result is
\begin{equation}
p(r) = \frac{1}{8} (8+r_0^2-r^2) \quad {\rm and} \quad q(r) = 2 .
\end{equation}
It is straightforward to verify that $p(r)$ and $q(r)$ satisfy
eqs.~(\ref{eq:polynomials}) for a disk profile. Finally, the full solution is
\begin{equation}
a_1(r_s)=\frac{k(r_s)}{\sqrt{r_0r_s}}\frac{(r_s^2-r_0^2)p(r_s)}{\pi r_s^2}
\, ,\quad
a_2(r_s)=\frac{\sqrt{r_0 r_s}}{k(r_s)}\frac{q(r_s)}{\pi r_s^2}\, ,\quad
a_3(r_s)=\frac{k(r_s)}{\sqrt{r_0r_s}}\frac{(r_0-r_s)^2(r_s^2+1)}{2 \pi r_s^2}\, .
\label{eq:full_disk_solution}
\end{equation}
When $r_s, r_0 \ll 1$ (in units of the Einstein radius)
these expressions can be further simplified. In this limit we also have
$k \approx \sqrt{n}$ and eq.~(\ref{eq:relPiE}) allow us to write
\begin{equation}
a_1(r_s) \approx \frac{2 (r_s - r_0)}{ \pi r_s^2}\, , \quad
a_2(r_s) \approx \frac{(r_s +r_0)}{\pi r_s^2}\quad {\rm and} \quad
a_3(r_s) \approx \frac{(r_s-r_0)^2}{\pi r_s^2 (r_0+r_s)}
\end{equation}
and for the total magnification one obtains
\begin{equation}
\mu_{\rm disk} (r_s, r_0) \approx
\frac{2 (r_s - r_0)}{ \pi r_s^2} K(\sqrt{n}) +
\frac{2 (r_s + r_0)}{ \pi r_s^2} E(\sqrt{n}) =
\frac{2 (1 - u)}{ \pi r_s} K\left(\frac{2\sqrt{u}}{1+u}\right) +
\frac{2 (1 +u)}{ \pi r_s} E\left(\frac{2\sqrt{u}}{1+u}\right)
\end{equation}
where $u=r_0/r_s$.
This result coincides with those presented in eqs.~(\ref{eq:I_ngeneral}, \ref{eq:efdisk})
with $n=0$. The derivation is described in Appendix~\ref{sec:series_expanion}.

Extended source effects start to become important when the projected position
of the lens intersects the source disk, i.e., when $r_0\simeq r_s$. Using the
properties of the elliptic integrals given in eq.~(\ref{eq:limEKPi}) we can write
\begin{equation}
\mu(r_s,r_s) = \frac{2}{\pi} \left[\frac{1}{r_s}+ \frac{1+r_s^2}{r_s^2}
\arctan(r_s) \right] \approx \frac{4}{\pi r_s} + \frac{4}{3 \pi} r_s -
\frac{4}{15 \pi} r_s^3 +...
\label{eq:b_rs_disc}
\end{equation}
Similar limits at $r_0\simeq 0$ and $r_0=r_s$ can be found for the other profiles
as we shall see below.

\subsection{Limb Darkening Profile}

For the limb darkening profile (eq.~\ref{eq:starprofiles2}) we have
\begin{equation}
c_3(r)= { 2 \over \pi r_s^2}((r^2-r_s^2) \sqrt{1- {r^2\over r_s^2}}+C_3)
\end{equation}
with $C_3$ a constant to be determined. Then,
\begin{equation}
a_3(r_s)=\frac{2C_3}{\pi r_s^2}\frac{(r_0-r_s)^2 k(r_s)}{4\sqrt{r_0r_s}}
\label{eq:a3_limb}
\end{equation}
From eq.~(\ref{eq:mu_limit}) we can compute the
amplification when the lens and source are perfectly aligned,
\begin{equation}
\label{mulimb0}
\mu_{\rm limb}(r_s, 0) = {3 \over r_s^2} \int_0^{r_s}
{(r^2+2)\over\sqrt{ 4+r^2 }}\sqrt{1-{r^2 \over r_s^2}} \, dr
= {2 \over k_2 r_s^2}\left[ (1+ {r_s^2 \over 2})
E(k_2) - K(k_2) \right] \approx \frac{3\pi}{4 r_s} + \frac{9\pi}{128} r_s
-\frac{15\pi}{4096} r_s^3+\cdots
\end{equation}
with $k_2 = r_s / \sqrt{4+ r_s^2}$. In this case, comparison with
eq.~(\ref{eq:a3_limb}) gives the limit $C_3=0$, that physically corresponds
to the source profile vanishing at the edge of the star.

Using the simplifying properties of elliptic integrals given in eq.~(\ref{eq:limEKPi})
for $r_0=r_s$ we obtain
\begin{equation}
\mu_{\rm limb}(r_s, r_s) \approx \frac{3\pi}{8 r_s} + \frac{135\pi}{1024} r_s
- \frac{1575 \pi}{2^{16}} r_s^3 + \cdots
\label{eq:b_rs_limb}
\end{equation}
Since most lensing events will have $r_s\ll 1$ (see the estimate given in
eq.~(\ref{eq:rs_magnitude})) this series converges very quickly.

\subsection{Parabolic Source Profile}

For the parabolic brightness profile of eq.~(\ref{eq:starprofiles3}) we have
\begin{equation}
c_3(r)= \frac{1}{\pi r_s^2}(2r^2(2-\frac{r^2}{r_s^2})+C_3)
\end{equation}
and
\begin{equation}
a_3(r_s)= \frac{(2 r_s^2+C_3)}{\pi r_s^2}\frac{(r_0-r_s)^2k(r_s)}{4\sqrt{r_0r_s}}
\end{equation}
The constant $C_3$ can be determined using the solution at $r_0=0$. In this
case the amplification is
\begin{equation} \label{mupara0}
\mu_{\rm para}(r_s, 0)=\frac{4}{r_s^2}\int_0^{r_s}
\frac{(r^2+2)}{\sqrt{4+r^2}}(1-\frac{r^2}{r_s^2})dr=
\frac{(2+r_s^2)\sqrt{4+r_s^2}}{r_s^3}-{8\over r_s^4}\arsinh({r_s\over 2})
\approx\frac{8}{3 r_s}+\frac{1}{5}r_s-\frac{1}{112} r_s^3 + \cdots
\end{equation}
that, like in the limb case, yields $C_3=0$.
Using the simplifying properties of elliptic function given by eq.~(\ref{eq:limEKPi})
when $r_0=r_s$, we can derive the following series expansion
\begin{equation}
\mu_{\rm para}(r_s, r_s) \approx \frac{32}{9 \pi r_s} + \frac{32}{25 \pi} r_s
- \frac{32}{147 \pi} r_s^3 + \cdots
\label{eq:b_rs_para}
\end{equation}
Again, as $r_s\ll 1$ this series converges very quickly.

\section{C. Series Expansion of the Single Lens Magnification}
\label{sec:series_expanion}

The integration of the angular part of eq.~(\ref{eq:In})
can be expressed in terms of elliptic integrals.
If we denote $k_1=2\sqrt{ut}/(u+t)$, the first three terms
in the series expansion become
\begin{eqnarray}
\nonumber
\hat{I}_0(u)&=&\int_{0}^{1}\frac{4}{(u+t)}K(k_1)\hat{s}(t)tdt, \\
\label{eq:I_n}
\hat{I}_1(u)&=&\int_{0}^{1} 4(u+t)E(k_1)\hat{s}(t)tdt,\\
\nonumber
\hat{I}_2(u)&=&\int_{0}^{1}\left[
\frac{4}{3}(u-t)(t^2-u^2)K(k_1) + \frac{16}{3}(u+t)(t^2+u^2)E(k_1)
\right]\hat{s}(t)tdt \, .
\end{eqnarray}
Similar combinations of complete elliptical integrals of the first, $K(k_1)$, and
second kind, $E(k_1)$, with higher order polynomials are obtained for $n>2$.
We will compute approximations up to $n=2$ since above this order the
contribution to the light curve is at most a 1\% (see Fig.~\ref{fig:fig3}).

The indefinite integrals can be evaluated assuming the result
has the form
\begin{equation}
I_n(t,u)=a_n(t,u)K(k_1)+b_n(t,u)E(k_1) .
\label{eq:I_n_ansatz}
\end{equation}
By differentiating eqs.~(\ref{eq:I_n}, \ref{eq:I_n_ansatz})
the coefficients verify the following differential equations
\begin{eqnarray}
\label{eq:dandt}
\frac{da_n}{dt}-(a_n+b_n)\frac{k_1^\prime}{k_1} &=& g_n(t,u) ,\\
\label{eq:dbndt}
\frac{db_n}{dt}+\left(\frac{a_n}{1-k_1^2}+b_n\right)\frac{k_1^\prime}{k_1}&=& h_n(t,u)
\end{eqnarray}
where the prime denotes the derivative with respect to the variable $t$, and
for $n=0,1,2$ we have
\begin{equation}
\begin{array}{ll}
g_0(t,u)=4t\hat{s}(t)/(t+u),  	& h_0(t,u)=0,\\
g_1(t,u)=0, 	 		& h_1(t,u)=4(u+t)t\hat{s}(t),\\
g_2(t,u)=\frac{4}{3}(u-t)(t^2-u^2) t\hat{s}(t),
& h_2(t,u)= \frac{16}{3}(u+t)(t^2+u^2)t\hat{s}(t).  \\
\end{array}
\end{equation}
Then, solving eqs.~(\ref{eq:dandt}, \ref{eq:dbndt}) for $n=0,1,2$ is equivalent to
integrate eqs.~(\ref{eq:I_n}). To simplify the differential equations of the
coefficients, we define the new functions $e_n, f_n$ as
$a_n(t,u)=(t-u)f_n(t)$ and $b_n(t,u)=(t+u)e_n(t)$. Using the identities
\begin{equation}
\frac{k_1^\prime}{k_1} = \frac{(u-t)}{2t (u+t)},  \quad
\frac{k_1^\prime}{(1-k_1^2)k_1} = \frac{(u+t)}{2t (u-t)} ,
\end{equation}
we obtain
\begin{eqnarray}
\label{eq:en_fn1}
&e_n&+2tf_n^\prime+\frac{3t^2+u^2}{t^2-u^2}f_n=\frac{2t}{t-u}g_n ,\\
\label{eq:en_fn2}
&e_n&+2te_n^\prime-f_n=\frac{2t}{t+u}h_n .
\end{eqnarray}
These equations can be combined to obtain a second order differential equation
for $e_n$
\begin{equation}
(t^2-u^2) e_n^{\prime\prime}+ \frac{(3t^2-u^2)}{t} e_n^\prime+e_n
= \frac{(t+u)}{2t} g_n + \frac{(3t-u)}{2t} h_n + (t-u) h_n^\prime \equiv Q_n(t)
= \sum_{m=0}^\infty q_{n,2m} t^{2m}
\label{eq:en}
\end{equation}
The r. h. s. of the differential equation $Q_n(t)$ results always
into a polynomial of even power as long as the source profile
$\hat{s}(t)$ has even power. We have
\begin{eqnarray}
Q_0(t) &=& 2 \hat{s}(t)\, \nonumber \\
Q_1(t) &=& (14t^2-6u^2)\hat{s}(t)+ 4(t^2-u^2) t \hat{s}^\prime(t)\, \\
Q_2(t) &=& \frac{2}{3}(43t^4+2t^2u^2-13u^4)\hat{s}(t)
+ \frac{16}{3} (t^4-u^4) t \hat{s}^\prime(t)
\nonumber
\end{eqnarray}
Now we are looking for solutions of $e(t)$ if $Q(t)$ is simply given by
$Q(t) = t^{2m}$.
This differential equation has trivial solutions of a form of a polynomial
$e(t) = P_{2m,u} (t) =\sum_{k=0}^m p_{2k} t^{2k}$ of the same power.
Inserting the polynomial into eq.~(\ref{eq:en}) gives
\begin{equation}
p_{2m} = \frac{1}{(2m+1)^2} \quad {\rm and} \quad p_{2k-2} = p_{2k} u^2
\frac{(2k)^2}{(2k-1)^2} \quad {\rm with}  \quad k = m, (m-1),...,1
\end{equation}
It is now easy to verify that $P_{2m,u}(t)$ obeys the recursion relation
\begin{equation}
P_{0,u}(t) = 1 \quad {\rm and} \quad
P_{2m,u}(t) =\frac{(2m)^2}{(2m+1)^2} u^2 P_{2m-2,u}(t) + \frac{t^{2m}}{(2m+1)^2}
\quad {\rm for}  \quad m= 1, 2,3,...
\end{equation}
with
\begin{equation}
P_{2,u}(t) = \frac{1}{9} ( t^2+4 u^2)\, , \quad
P_{4,u}(t) = \frac{1}{225} (9 t^4 +16 t^2 u^2 + 64 u^4),
\end{equation}
and so on. In general we can now construct the solution $e_n(t)$ for any given
$Q_n(t)$. The solution has the form
\begin{equation}
e_n(t) = \sum_{m=0}^\infty q_{n,2m} P_{2m, u} (t)
\end{equation}
In summary, $f_n(t)$ and $e_n(t)$ are polynomials of degree $2n$ in $t$ and $u$
for a disk and of degree $2n+2$ for a parabolic profile, respectively.
If we expand the limb darkening profile until degree 10 the resulting
$f_0$ and $e_0$ have powers of 10 in $t$ and $u$ respectively.
Each higher order increase the power by 2.

Since we have constructed now a solution $e_n(t)$ and $f_n(t)$
we are able to write down for each source profile a set of solutions.
For example, for $I_0$ we can write
\begin{equation}
\hat{I}_0(u)=
\int_{0}^{1} \frac{4}{(u+t)} K ( k_1 ) \hat{s}(t) t dt =
\big[ (t-u)[e_0(t)+2te_0^\prime(t)]K(k_1)
+(t+u)e_0(t)E(k_1) \big]\bigg|_0^1 .
\end{equation}
Let us remark that the lower limit of the integral always vanishes so
we need to pay attention only to the upper limit. For each source profile
\begin{eqnarray}
\hat{s}_{\rm disk}(t) &=& 1 \hspace*{118pt}
 \Rightarrow \quad e_0(t) = 2 P_{0,u}(t), \nonumber \\
\hat{s}_{\rm para}(t) &=& 2(1-t^2) \hspace*{86pt}
\Rightarrow\quad e_0(t)=4(P_{0,u}(t)-P_{2,u}(t)),\\
\hat{s}_{\rm limb}(t) &=&\frac{3}{2}\sqrt{1-t^2} =\frac{3}{2}
\sum_{k=0}^{\infty} {\frac{1}{2} \choose k} t^{2k}
\quad\Rightarrow\quad
e_0(t)=3 \sum_{k=0}^{\infty}{\frac{1}{2}\choose k} P_{2k,u}(t) .
\nonumber
\end{eqnarray}
Evaluating these expressions at the surface of the star, $t=1$, we obtain the
results given in the main text. We note that the integral has the form
\begin{equation}
\hat{I}_n(u)= (1-u) f_n(1) K\left(\frac{2\sqrt{u}}{1+u}\right)
+(1+u)e_n(1) E\left(\frac{2\sqrt{u}}{1+u}\right) ,
\end{equation}
with
\begin{equation}
f_0(t) = e_0(t) + 2t e_0^\prime (t) \, , \quad
f_1(t) = e_1(t) + 2t e_1^\prime (t) -8 t^2 \hat{s}(t), \quad
f_2(t) = e_2(t) + 2t e_2^\prime (t) - \frac{32}{3} (t^2+u^2) t^2 \hat{s}(t) ,
\end{equation}
Note that in the main text we have defined
$e_n(u) \equiv e_n(t=1)$ and $f_n(u) \equiv f_n(t=1)$.

\section{D. Functional Relations of Complete Elliptic Integrals}
\label{sec:ellipticfct}

The following properties of the first, second and third complete elliptical
integrals (see \citep{BF71})
\begin{eqnarray}
{dK(k) \over dk} &=& { E(k) \over k(1-k^2) } - { K(k) \over k} \\
{dE(k) \over dk} &=& { E(k) -K(k) \over k } \\
{\partial \Pi(n,k) \over \partial n } &=& { E(k) \over 2 (k^2-n) (n-1) }
+{K(k) \over 2 n (n-1) }+{(n^2-k^2) \Pi(n,k) \over 2 (k^2-n) n (n-1) } \\
{\partial \Pi(n,k) \over \partial k } &=& { k E(k) \over (n-k^2) (k^2-1) }
+{k\Pi(n,k) \over (n-k^2)  } .
\end{eqnarray}
where used to derive eqs.~(\ref{eq:a1}), (\ref{eq:a2}) and (\ref{eq:a3}).
Also, the following relations
\begin{equation} \label{eq:intEK}
\int { K(k) \over k^2 } dk = - {E(k) \over k} , \quad
\int { E(k) \over 1-k^2 } dk = k K(k)
\end{equation}
and the Legendre relation
\begin{equation} \label{eq:Legendre}
E(k) K(\hat{k}) +E(\hat{k}) K(k)- K(k) K(\hat{k}) = {\pi \over 2}
\end{equation}
with $\hat{k} = \sqrt{1-k^2}$, were also used.
In the limiting case with $r\rightarrow r_0$ we have the following
relations
\begin{equation} \label{eq:limEKPi}
\lim_{r\rightarrow r_0} (r-r_0) K(k) = 0, \quad
\lim_{r\rightarrow r_0} E(k) = E(1) = 1
\quad {\rm and} \quad
\lim_{r\rightarrow r_0} (r-r_0)^2\Pi(n,k) = 4 r_0 \arctan (r_0)
\end{equation}
where we used the definition of $n$ and $k$ of eq.(\ref{eq:nk}).
For small values $r,r_0 \ll 1$ we have $k \approx \sqrt{n}$ so that
we can make use of the relation
\begin{equation} \label{eq:relPiE}
\Pi(n, \sqrt{n} ) = \frac{E(\sqrt{n}) }{(1-n)} = \frac{(r_0 + r)^2}{(r_0-r)^2}
E(\sqrt{n})
\end{equation}

\end{document}